\documentclass[sort&compress]{elsarticle}
\usepackage{lineno}
\usepackage{graphicx}
\usepackage{dcolumn}
\usepackage{bm}
\usepackage{upgreek}
\usepackage{hyperref}
\usepackage{amsmath}
\usepackage{amssymb}
\usepackage{multirow}
\usepackage[margin=1.3in]{geometry}
\let\oldequation\equation
\let\oldendequation\endequation
\renewenvironment{equation}
  {\linenomathNonumbers\oldequation}
  {\oldendequation\endlinenomath}
\usepackage{mdframed} 
\usepackage{multicol} 
\usepackage{nomencl} 
\makenomenclature
\usepackage{hyperref}
\usepackage{soul,color,xcolor}
\soulregister{\citep}7
\soulregister{\citet}7
\soulregister{\ref}7
\soulregister{\pageref}7
\usepackage{etoolbox}
\renewcommand\nomgroup[1]{%
  \item[\bfseries
  \ifstrequal{#1}{G}{Greek symbols}{%
  \ifstrequal{#1}{S}{Subscripts}{}}%
]}

\newcommand{\We}{W\!e}%
\newcommand{\Oh}{Oh}%
\journal{Journal of Colloid and Interface Science}
\begin{document}
\begin{frontmatter}
\title{Deformation and breakup of compound droplets in airflow}
\author[SKLE]{Zhikun Xu}
\author[SKLE]{Yue Zhang}
\author[SKLE,PES]{Tianyou Wang}
\author[SKLE,PES]{Zhizhao Che\corref{cor1}}
\cortext[cor1]{Corresponding author.
}
\ead{chezhizhao@tju.edu.cn}
\address[SKLE]{State Key Laboratory of Engines, Tianjin University, Tianjin, 300350, China.}
\address[PES]{National Industry-Education Platform of Energy Storage, Tianjin University, Tianjin, 300350, China}
\begin{abstract}
\emph{Hypothesis}: Immiscible liquids are commonly used to achieve unique functions in many applications, where the breakup of compound droplets in airflow is an important process. Due to the existence of the liquid-liquid interface, compound droplets are expected to form different deformation and breakup morphologies compared with single-component droplets. \\
\emph{Experiments}: We investigate experimentally the deformation and breakup of compound droplets in airflow. The deformation characteristics of compound droplets are quantitatively analyzed and compared with single-component droplets. Theoretical models are proposed to analyze the transition between breakup morphologies. \\
\emph{Findings}: The breakup modes of compound droplets are classified into shell retraction, shell breakup, and core-shell breakup based on the location where the breakup occurs. The comparison with single-component droplets reveals that the compound droplet is stretched more in the flow direction and expands less in the cross-flow direction, and these differences occur when the core of the compound droplet protrudes into the airflow. The transition conditions between different breakup modes are obtained theoretically. In addition, the eccentricity of the compound droplet can lead to the formation of the thick ligament or the two stamens in the droplet middle.
\end{abstract}
\begin{keyword}
\texttt {
Compound droplet \sep
Droplet breakup \sep
Droplet deformation \sep
Secondary breakup \sep
Atomization
}
\end{keyword}
\end{frontmatter}
\def \scaleSize {0.8}
\def \scaleSiz2 {0.6}
\begin{table}[]
\label{tab:my-table}
\begin{tabular}{|llll|}
\hline
\multicolumn{4}{|l|}{\textbf{Nomenclature}}\\
$a$         & acceleration                      & $u_g$                             & velocity of the airflow              \\
$A_p$       & projected area of the oil shell   & $u_L$                             & velocity of the leftmost point       \\
$C$         & constant                          & $V_c$                             & volume flow rate of the inner needle \\
$d_0$       & diameter of the compound droplet  & $V_0$                             & total volume flow rate               \\
$d_c$       & diameter of the water core        & \multicolumn{2}{l|}{\textit{Greek letters}}                              \\
$d_{c,ex}$  & experimental diameter of the core & $\alpha $                         & volume ratio of the water core       \\
$d_{c,th}$  & theoretical diameter of the core  & $\varepsilon $                    & eccentricity of the water core       \\
$d_{c,w}$   & water core width                  & ${\lambda _{RT}}$                 & RT instability wavelength            \\
$d_{os}$      & oil shell width on one side       & ${\mu _o}$                        & oil viscosity                        \\
$d_{o,w}$   & oil shell width                   & ${\rho _g}$                       & air density                      \\
$d_t$       & thickness of the compound droplet & ${\rho _o}$                       & oil density                          \\
$d_w$       & width of the compound droplet     & ${\rho _w}$                       & water density                        \\
$F_D$       & airflow drag force                & ${\sigma _o}$                     & oil surface tension                  \\
$h$         & thickness of the oil shell        & ${\sigma _{ow}}$                  & oil-water interfacial tension        \\
$H$         & droplet falling height            & ${\sigma _w}$                     & water surface tension                \\
$m$         & mass of the oil shell             & \multicolumn{2}{l|}{\textit{Dimensionless numbers}}                      \\
$s$         & distance                          & $Oh$                              & Ohnesorge number                     \\
$t$         & time                              & $\We_g$                             & Weber number of the oil shell        \\
${t^ * }$   & characteristic time               & ${\We_{g,wc}}$                     & Weber number of the water core       \\
\hline
\end{tabular}%
\end{table}

\section{Introduction}\label{sec:1}
The breakup of droplets in airflow is encountered in numerous applications, including fuel atomization \cite{Mukhtar2019TrifuelEmulsion}, agricultural spray \cite{Li2021EmulsionSprays, Iaroslav2021SheetBreakup}, food processing \cite{Selvam2019SprayDryingFood, Fieber2011Encapsulation}, and drug encapsulation and delivery \cite{Groneberg2003DrugDelivery, Xie2007Electrospray, Yin2009Encapsulation}. In these applications, immiscible liquids are commonly used to achieve unique functions, improve fluid atomization, or achieve controlled release, which leads to the formation of compound droplets. Different breakup behaviors of compound droplets are expected for different applications. For instance, in agricultural sprays, the breakup of compound droplets is necessary to enhance the coverage of active ingredients on crops \cite{Li2021EmulsionSprays}, while in drug delivery, the breakup of compound droplets should be minimized to avoid reduction of efficacy or loss of functionality \cite{Groneberg2003DrugDelivery}. Therefore, understanding the breakup of compound droplets is crucial for designing and optimizing these processes.

The breakup of single-component droplets caused by airflow has been investigated in many studies \cite{Sharma2022BreakupReview, Guildenbecher2009SecondaryAtomization}. The Weber number ($\We_g$) and the Ohnesorge number ($\Oh$) are the main controlling parameters, representing the magnitude of the aerodynamic force of the airflow and the viscous force of the droplet relative to its surface tension, respectively \cite{Hsiang1992SecondaryBreakup}. At the same $\Oh$, as $\We_g$ increases, the droplet may exhibit different breakup modes sequentially, including oscillatory deformation, bag, bag-stamen, multimode, and shear-stripping breakup \cite{Sharma2022BreakupReview, Xu2023Viscosity}. The different breakup modes are mainly controlled by the competition of the Rayleigh-Taylor (RT) instability and the Kelvin-Helmholtz (KH) instability, corresponding to the piercing induced by the RT instability at relatively lower $\We_g$ and the stripping by the KH instability at relatively higher $\We_g$ \cite{Theofanous2011DropBreakup, Guildenbecher2009SecondaryAtomization, Sharma2022BreakupReview}. Under different breakup modes, different droplet morphological and dynamic characteristics can be observed, which further results in different breakup outcomes, such as different size distributions of fragments \cite{Someshwar2023SizeDistribution, Jackiw2022SizeDistribution}, and different deformation and breakup times \cite{Guildenbecher2009SecondaryAtomization, Theofanous2011DropBreakup}. For instance, a droplet in bag breakup mode shows a morphology with a thin bag film surrounded by a thick liquid rim. After the bag breakup of droplets, the total fragments are composed of fragments of the bag film, the edge ring, and the edge node, exhibiting a tri-modal size distribution \cite{Someshwar2023SizeDistribution, Jackiw2022SizeDistribution}. In addition, non-Newtonian properties also affect the breakup of droplets \cite{Che2023ShearThinningBreakup, Mitkin2017StrainthickeningDroplet, Sharma2022BreakupReview, Theofanous2013ViscoelasticDroplet}. The deformation and breakup of non-Newtonian droplets exhibit some new features, such as the apparent persistent ligaments of viscoelastic droplets \cite{Theofanous2013ViscoelasticDroplet}. Even unique breakup modes may emerge, such as the hemline breakup mode of gel droplets \cite{Wang2023GelDrop}.

Compared with single-component droplets, compound droplets consist of two (or more) immiscible liquids with different properties, such as different interfacial tension and different viscosity, and these different properties further affect the breakup of droplets. The breakup of compound droplets has been studied mainly for impingements \cite{Nathan2020CompoundDropletRebound, Blanken2021CompoundDropletImpact, Damak2022ImpactingEmulsionDroplet, Yin2022TripleLayeredEncapsulation}, microfluidics \cite{Liu2021LiquidLiquidShear, Nabavi2015microfluidic}, and puffing/micro-explosions \cite{Antonov2019ExplosionEmulsions, Wang2022Microexplosion}, but these processes are fundamentally different from the aerodynamic breakup of droplets in airflow. In addition, the aerodynamic breakup of droplets containing solid particles \cite{Zhao2021ShearThickening, Zhao2011Slurry} or a vapor cavity \cite{Liang2020DropletCavity} has been studied. Zhao et al.\ \cite{Zhao2021ShearThickening, Zhao2011Slurry} studied the breakup of droplets containing solid particles in airflows, and found the presence of solid particles makes the droplet exhibit some non-Newtonian properties and further leads to the formation of new breakup modes. Liang et al.\ \cite{Liang2020DropletCavity} investigated the interaction of a shock wave and a droplet with a vapor cavity, and found that the cavity first collapses and then expands, and the cavity evolution affects the deformation of the droplet's outer surface. These studies have demonstrated that multi-component droplets can result in novel aerodynamic breakup phenomena. However, there is a lack of research on the aerodynamic breakup of compound droplets composed of two immiscible liquids.

In this study, the aerodynamic breakup of compound droplets is studied by producing compound droplets composed of water cores and silicone oil shells (Section \ref{sec:2}). The breakup morphologies (Section \ref{sec:31}) and deformation characteristics (Section \ref{sec:32}) of compound droplets are discussed and compared with those of single-component silicone oil droplets. Moreover, the regime map of compound droplet breakup is presented, and the transitions between different breakup regimes are obtained through theoretical analysis (Section \ref{sec:33}). Finally, the effect of eccentricity is discussed (Section \ref{sec:34}).

\section{Experimental method}\label{sec:2}
\begin{figure}
  \centering
  \includegraphics[width=\columnwidth]{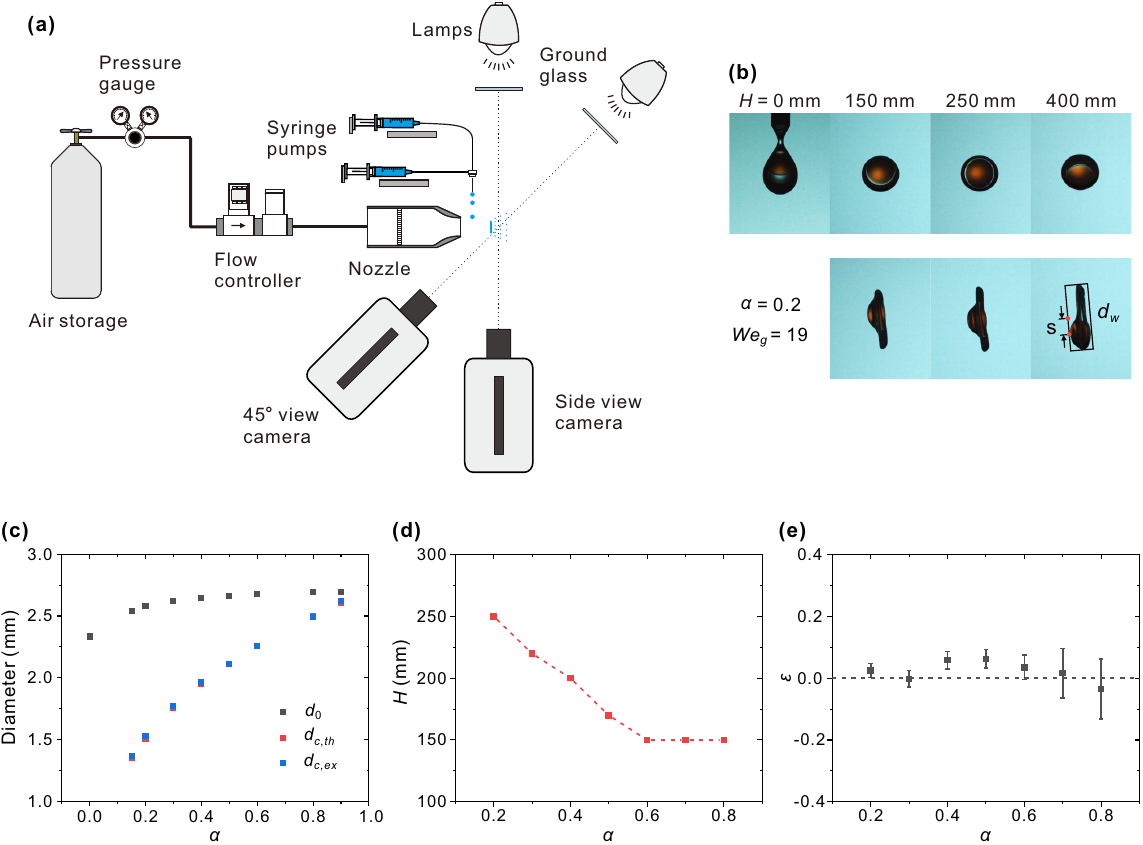}
  \caption{(a) Experimental setup for droplet breakup. (b) Positions of the water core at different falling heights. The first row shows the images without airflow and the second row shows the images after the droplets are flattened in the airflow. (c) Diameters of compound droplets with different volume ratios of water core ($\alpha$). (d) Falling height of the droplet at different $\alpha$. (e) Eccentricities and the corresponding uncertainties at the falling heights in (d). The corresponding $\We_g$ increases from 30 to 60 with increasing $\alpha$ to enable a clear presentation of the water core position.}\label{fig:01}
\end{figure}

The experimental setup for the breakup of compound droplets is shown in Figure \ref{fig:01}a. A continuous jet was used to study droplet breakup, which was generated in the same way as our previous studies \cite{Xu2022ShearBreakup, Xu2023Viscosity}. The airflow was formed from compressed air storage, adjusted by a mass flow controller, maintained by a long pipe (length 600 mm, containing a honeycomb and two-layer mashes), and finally ejected from a converging nozzle (smoothly reducing the cross-section from 40 × 40 to 20 × 30 mm$^2$). The airflow velocities were $u_g =$ 8.9--70 m/s, and the density of the airflow was $\rho_g = 1.2$ kg/m$^3$ at an experiment temperature of 20 $^\circ$C. The velocity field profile of the airflow is similar to that measured in our previous study \cite{Xu2022ShearBreakup}, which consists of a core region, a shear layer, and an outer region. The range of the core region of the jet is about a height of 30 mm and a length of $ > 50$ mm. The compound droplet entered the airﬂow from above by gravity. The falling velocity of the droplet (1.7--2.2 m/s)   is large enough to ensure that the droplet passes through the shear layer quickly and the breakup is mainly in the core region of the jet. Two cameras were used to capture the breakup process of compound droplets in the airflow. The camera for the side view (Phantom v1612) had a frame rate of 10000 fps and a resolution of 45 $\mu$m/pixel. The camera for the 45$^\circ$ view (Photron Fastcam SA1.1) had a frame rate of 5000 fps and a resolution of 56 $\mu$m/pixel. In addition, the background light sources were two 280-W LED lamps diffused by ground glasses.

The compound droplet used in this study is composed of a core of water and a shell of silicone oil. The silicone oil had a viscosity ${\mu _o} = 10$ mPa$\cdot$s, a density $\rho_o = 930$ kg/m$^3$, and a surface tension $\sigma_o = 20$ mN/m. The deionized water had a density $\rho_w = 998$ kg/m$^3$, and a surface tension $\sigma_w = 72$ mN/m. The water was dyed red by adding Congo red (less than 0.1 wt\%) to enhance the contrast with the oil. The oil-water interfacial tension ($\sigma_{ow}$) was $\sim$40 mN/m according to previous studies \cite{Nathan2020CompoundDropletRebound, Yang2023SingularJet}. The compound droplet was generated through a coaxial needle \cite{Nathan2020CompoundDropletRebound, Blanken2021CompoundDropletImpact}. Before the experiment, the coaxial needle was hydrophobized by placing it in a 0.5\% anti-fingerprint (AF) oil solution for 10 min and then dried in an oven (120 $^\circ$C for 40 min). This is to avoid the water inside the inner needle blocking the oil flow through the outer needle \cite{Nathan2020CompoundDropletRebound}. After the needle was hydrophobized, the silicone oil and the water were infused through the outer needle (gauge 18) and the inner needle (gauge 25), respectively. The flow rates of the two liquids were adjusted by two syringe pumps (Harvard Apparatus, Pump 11 Elite), and the total flow rate was fixed at 40 ml/h. By varying the flow rates of the two liquids, different volume ratios of the water core ($\alpha$) in the compound droplet were obtained
\begin{equation}\label{eq:alpha}
  \alpha=\frac{V_c}{V_0}=\frac{d_c^3}{d_0^3}
\end{equation}
where $V_c$ and $d_c$ are the volume and diameter of the water core, and $V_0$ and $d_0$ are those of the compound droplet, respectively. The volume ratio ($\alpha$) is calculated as the ratio of the flow rates of the inner and outer needles. To verify the validity of the coaxial needle method, we measured the diameters of the compound droplet ($d_0$) and the water core (${{d}_{c,ex}}$). $d_0$ was measured directly in the air, and a theoretical diameter of the water core (${{d}_{c,th}}$) was calculated based on $d_0$, i.e., ${{d}_{c,th}}={{d}_{0}}{{\alpha }^{1/3}}$. The real diameter of the water core (${{d}_{c,ex}}$) was obtained by dropping the compound droplet into a liquid pool consisting of the same silicone oil, and then measuring the core diameter. As shown in Figure \ref{fig:01}c, ${{d}_{c,ex}}$ was almost equal to ${{d}_{c,th}}$ at different $\alpha$, indicating that the coaxial needle method to generate compound droplets is robust.

With the coaxial needle method, the water core is at the uppermost position in the oil shell when the compound droplet is just formed. As the droplet falls, the water core moves downwards in the oil shell. Therefore, the eccentricity of the water core can be tuned via the droplet falling height ($H$), as shown in Figure \ref{fig:01}b. However, the direct measurement of the eccentricity is difficult due to the thin thickness of the oil shell and the light refraction caused by the oil shell (as shown in the first row of Figure \ref{fig:01}b). Considering that the relative position of the water core can be better presented after the droplets are flattened in the airflow (as shown in the second row of Figure \ref{fig:01}b, the protruding part is the water core, and the flattened part is the oil shell), we measure the eccentricity of the water core ($\varepsilon$) after the droplets are flattened in the airflow. Since the speed of the droplet flattening is much faster than the speed of the water core relative to the shell, the eccentricity after flattening can be regarded as the same as the eccentricity before the flattening. As shown in Figure \ref{fig:01}b, we first find the smallest bounding box of the droplet using image processing. Then, the eccentricity is calculated as $\varepsilon = s/\left( {{d_w}/2} \right)$, where $s$ is the distance between the left midpoint of the smallest bounding box and the left intersection of the bounding box and the droplet profile (as illustrated in Figure \ref{fig:01}b), and $d_w$is the width of the droplet. In the present study, we mainly consider the scenarios where the eccentricity is 0, that is the water core is at the center of the oil shell. The corresponding falling height of the droplet at different $\alpha$ was shown in Figure \ref{fig:01}d. Figure \ref{fig:01}e shows the eccentricities and the corresponding uncertainties at the falling height given in Figure \ref{fig:01}d. It can be seen that the eccentricities are overall very small ($\varepsilon < 0.1$) at these falling heights. In addition, the uncertainties increase at larger volume ratios ($\alpha $). This is because as $\alpha $ increases, the water core deforms more, which makes finding the left intersection of the bounding box and the droplet profile unstable in the image processing.

\section{Results and discussion}\label{sec:3}
\subsection{Breakup morphologies of compound droplets}\label{sec:31}
The compound droplet can exhibit a complex breakup morphology due to the different deformation rates of the oil shell and the water core. Here, according to whether and where the breakup occurs, we classify the breakup of compound droplets into shell retraction (where the oil shell is flattened and then retracts), shell breakup (where the oil shell breaks up), and core-shell breakup (where both the oil shell and the water core break up). The main control parameters are the volume ratio of the water core ($\alpha$) and the Weber number (${{\We}_{g}}={{\rho }_{g}}{{d}_{0}}u_{g}^{2}/{{\sigma }_{o}}$). In addition, the viscosity of the droplet can be negligible in this study due to  $Oh \approx 0.045 < 0.1$ \cite{Sharma2022BreakupReview, Xu2023Viscosity}, where the Ohnesorge number is defined as $Oh = {\mu _o}/\sqrt {\left[ {\alpha {\rho _w} + \left( {1 - \alpha } \right){\rho _o}} \right]{d_0}{\sigma _o}} $. In this inviscid case, the breakup processes of single-component oil and water droplets under the same $\We_g$ should be similar. Hence, we only compare the differences between compound droplets and single-component oil droplets at similar $\We_g$. Moreover, the time is nondimensionalized by ${t^ * } = {d_0}\sqrt {\left[ {\alpha {\rho _w} + \left( {1 - \alpha } \right){\rho _o}} \right]/{\rho _g}} /{u_g}$ for comparison, which is defined as the time required for a droplet to move a characteristic length $d_0$ at a velocity ${u_d} = {u_g}\sqrt {{\rho _g}/\left[ {\alpha {\rho _w} + \left( {1 - \alpha } \right){\rho _o}} \right]} $ caused by the aerodynamic pressure \cite{Jain2019HighDensity, Ranger1969Aerodynamic}.

\subsubsection{Shell retraction}\label{sec:311}
\begin{figure}
  \centering
  \includegraphics[width=0.8\columnwidth]{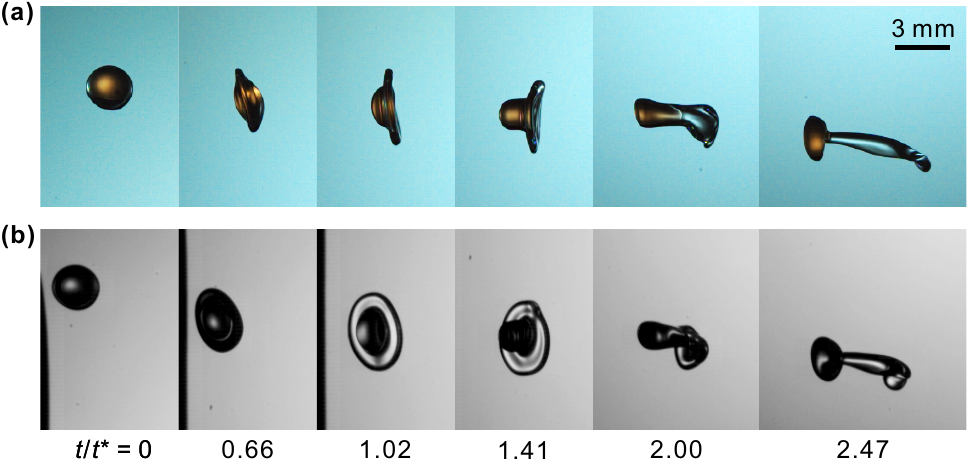}
  \caption{Shell retraction of a compound droplet at $\We_g = 21.5$, $\alpha = 0.3$, (a) side view, (b) 45$^\circ$ view. The whole process is shown as Video 1 in Supplementary Material.}\label{fig:02}
\end{figure}

A typical shell retraction process of a compound droplet is shown in Figure \ref{fig:02}. When the compound droplet enters the airflow, the oil shell is quickly flattened, while the water core is less deformed due to its larger surface tension ($t/t^* = 1.02$ in Figure \ref{fig:02}). After the flattening of the oil shell, the peripheral oil shell moves faster along the airflow due to the larger aerodynamic drag caused by the larger windward area of the oil shell ($t/t^* = 1.41$ in Figure \ref{fig:02}). Then the peripheral oil shell retracts behind the middle water core ($t/t^* = 2.00$ in Figure \ref{fig:02}). The droplet is not pierced by the airflow during this process, but it may break into a few large fragments as the droplet oscillates in the final stage ($t/t^* = 2.47$ in Figure \ref{fig:02}).

\subsubsection{Shell breakup}\label{sec:312}
\begin{figure}
  \centering
  \includegraphics[width=0.8\columnwidth]{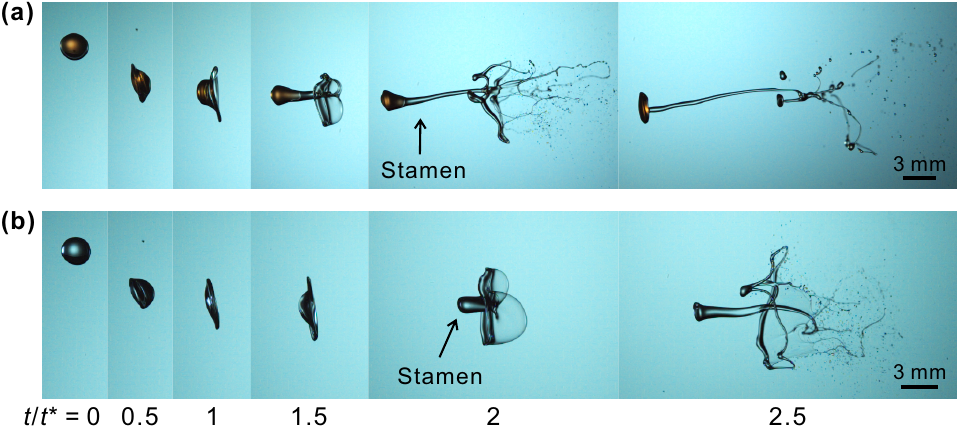}
  \caption{(a) Shell breakup of a compound droplet at $\We_g = 30.4$, $\alpha = 0.3$, (b) bag-stamen breakup of a single-component droplet at $\We_g = 27.2$, $\alpha = 0$. The whole process is shown as Video 2 in Supplementary Material.}\label{fig:03}
\end{figure}

A typical shell breakup process of a compound droplet is shown in Figure \ref{fig:03}a. For the shell breakup of the compound droplet, the oil shell deforms quickly and is subsequently pierced to form a bag film, while the water core undergoes oscillatory deformation but does not break up. The water core inhibits the deformation of the middle of the compound droplet, so the compound droplet does not undergo the classic bag breakup, where the droplet is pierced from the droplet middle \cite{Sharma2022BreakupReview}. In contrast, the shape of the compound droplet is similar to the classical bag-stamen breakup of a single-component droplet. Hence, a bag-stamen breakup of a single-component silicone oil droplet under a similar condition is shown in Figure \ref{fig:03}b for comparison. From the comparison, we can see some different features. Compared with the middle stamen of the classical bag-stamen breakup, the middle stamen of the shell breakup is composed of a head of the water core and a tail of the oil ligament. In addition, the stamen of the shell breakup appears earlier ($t/t^* = 1$ in Figure \ref{fig:03}a) than the single-component droplet, and the ligament of the stamen is stretched more significantly ($t/t^* = 2$ in Figure \ref{fig:03}a). Further quantitative comparisons of the deformation will be discussed in Section \ref{sec:32}.

\subsubsection{Core-shell breakup}\label{sec:313}
\begin{figure}
  \centering
  \includegraphics[width=0.85\columnwidth]{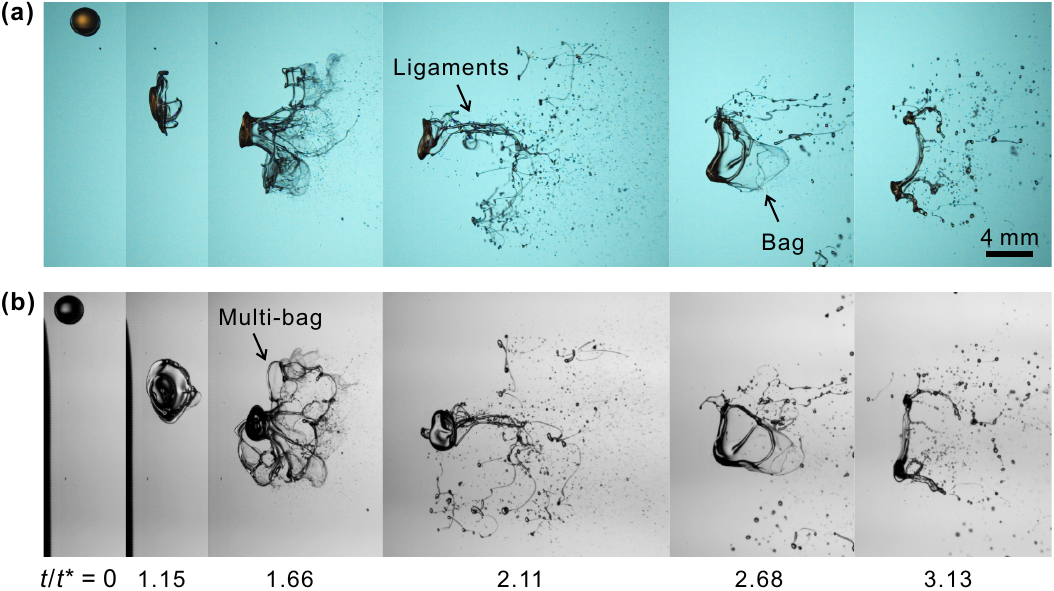}
  \caption{Core-shell breakup of a compound droplet at $\We_g = 88$, $\alpha = 0.3$, (a) side view, (b) 45$^\circ$ view. The whole process is shown as Video 3 in Supplementary Material.}\label{fig:04}
\end{figure}

A typical core-shell breakup process of a compound droplet is shown in Figure \ref{fig:04}. For the core-shell breakup, both the oil shell and the water core of the compound droplet break up, but at different time. The oil shell of the compound droplet is first pierced to form a multi-bag structure ($t/t^* = 1.66$ in Figure \ref{fig:04}). Then, the water core of the droplet forms a bag structure and breaks up ($t/t^* = 2.68$ in Figure \ref{fig:04}). Due to the different deformation rates of the peripheral oil shell and middle water core, distinct ligaments are stretched at their junction ($t/t^* = 2.11$ in Figure \ref{fig:04}). Moreover, as $\We_g$ increases further, more complex breakup structures may appear in the water core. For instance, after the peripheral oil shell is broken, the middle water core presents a multimode breakup structure (as shown in Video 4 in Supplementary Material).

\subsection{Deformation characteristics of compound droplets}\label{sec:32}
As discussed in Section \ref{sec:31}, the presence of water cores in compound droplets can affect their deformation and breakup morphologies. In this section, the droplet deformation is further quantified by droplet width ($d_w / d_0$), droplet thickness ($d_t / d_0$), and their ratio ($d_w / d_t$). The evolutions of these deformation characteristics of the compound droplet with time and $\We_g$ are obtained through image processing. The image processing includes removing the background, binarizing the image, and finding the smallest bounding box of the droplet. The details of image processing can be found in our previous study \cite{Xu2023Viscosity}. In addition, to improve the measurement uncertainty, each plot in this section shows the statistics of about 6 repeated experiments.

\begin{figure}
  \centering
  \includegraphics[width=0.85\columnwidth]{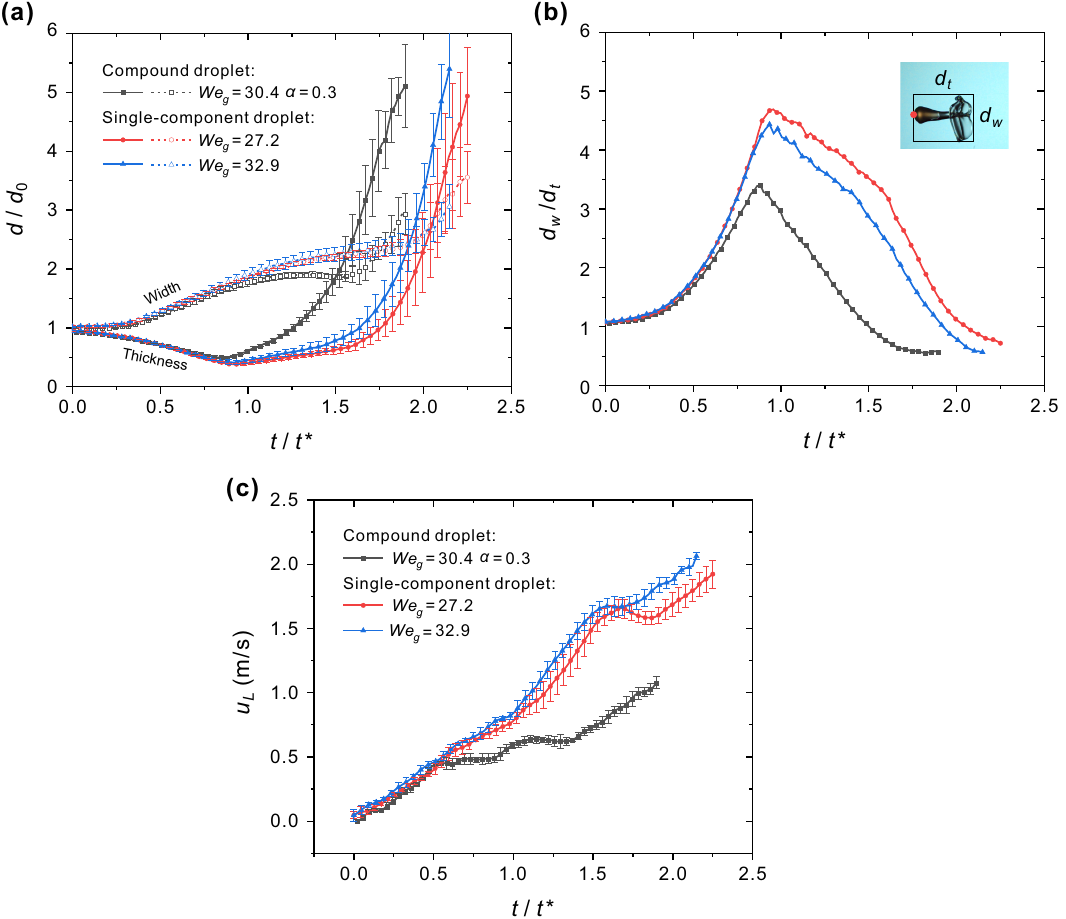}
  \caption{(a) Evolution of the width ($d_w / d_0$, the dashed lines) and the thickness ($d_t / d_0$, the solid lines) of the droplets with time. (b) Evolution of the ratio ($d_w / d_t$) of the width to the thickness with time. (c) Evolution of the velocity of the leftmost point on the droplet windward side with time.}\label{fig:05}
\end{figure}

Figures \ref{fig:05}a-b show the shape evolution of the compound droplet with time and compare it with that of the single-component silicone oil droplet. The deformation of the droplets in airflow can be divided into two stages, including the initial flattening stage (about $t/t^* =$ 0--1) and the later bag development stage (about $t/t^* > 1$). The transition of the two stages corresponds to the moment when the aspect ratio ($d_w / d_t$) is the largest, that is the peak in Figure \ref{fig:05}b. In the flattening stage, the droplet thickness ($d_t / d_0$) decreases, and the droplet width ($d_w / d_0$) increases. In the bag development stage, both the thickness and width of the droplet increase. Compared with the single-component droplet at similar $\We_g$, the thickness of the compound droplet is larger and grows faster, and the width is smaller but also grows faster. This indicates that the compound droplet is stretched along the airflow direction earlier and faster. In addition, in the bag development stage (about $t/{t^*} > 1$), the ratio ${d_w}/{d_t}$ at ${\We_g} = $ 32.9 is smaller than that at ${\We_g} = $ 27.2, as shown in Fig.~\ref{fig:05}b. This is because the size of the thin bag film is sensitive to the variation of ${\We_g}$ in the bag development stage. Hence, the thickness $d_t$ (determined by the size of the bag film) increases faster at ${d_w}/{d_t}$ at ${\We_g} = $ 32.9 than at ${\We_g} = $ 27.2 ($t/{t^*}$ = 1--2 of solid lines in Figure \ref{fig:05}a), while the width $d_w$ is similar at both $\We_g$ ($t/{t^*}$ = 1--2 of dashed lines in Figure \ref{fig:05}a). The large $d_t$ causes the ratio ${d_w}/{d_t}$ at ${\We_g} = $ 32.9 to be smaller than that at ${\We_g} = $ 27.2.

To further explore the reason for the different shape evolution, we extract the velocity ($u_L$) of the leftmost point of the droplet (the red point indicated in the inset of Figure \ref{fig:05}b), as shown in Figure \ref{fig:05}c. For the single-component silicone oil droplet, the velocity of the leftmost point increases steadily. However, for the compound droplet, the increase in velocity of the leftmost point slows down at $t/t^* \sim 0.5$. Combined with the high-speed images in Figure \ref{fig:03}a, $t/t^* \sim 0.5$ corresponds to the instant when the water core starts to protrude into the airflow, i.e., the velocity of the leftmost point starts to be dominated by the motion of the water core instead of the oil shell. Compared with the oil shell, the deformation of the water core is slower due to the larger surface tension of the water core, and it further leads to a slower increase in the velocity of the leftmost point. Moreover, the stretching of the droplet is controlled by the velocity difference between the periphery and the middle of the droplet. The velocity of the periphery of the single-component droplet and the compound droplet is similar due to the same component (oil shell) and the similar $\We_g$. The velocity of the droplet middle is indicated by the velocity of the leftmost point. Hence, the velocity difference between the periphery and the middle of the compound droplet is larger since the velocity of the leftmost point of the compound droplet is smaller, which indicates a faster stretching of the compound droplet. The difference in the velocity of the leftmost point between the single-component droplet and the compound droplet starts at $t/{t^ * } \sim 0.5$, as shown in Figure \ref{fig:05}c. The resulting difference in the droplet shape develops gradually and becomes apparent at a later stage, as shown in Figures \ref{fig:05}a-b.

\begin{figure}
  \centering
  \includegraphics[width=0.9\columnwidth]{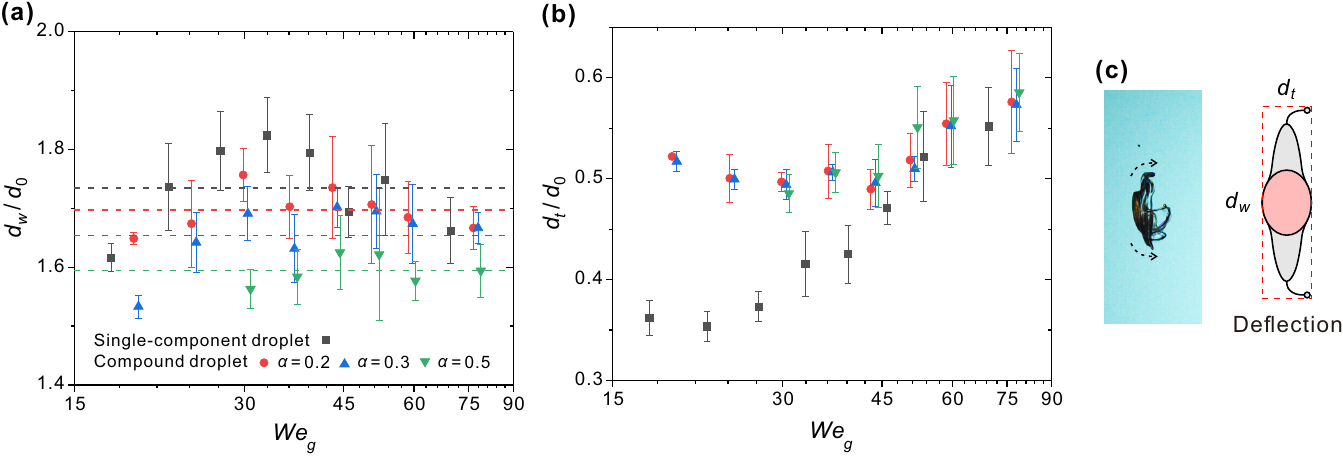}
  \caption{Variation of (a) droplet width and (b) droplet thickness with $\We_g$ at the instant when the droplet reaches the largest aspect ratio (corresponding to the peak in Figure \ref{fig:05}b). The dashed lines correspond to the average droplet width of different $\We_g$. (c) Illustration of the droplet deflection at a high $\We_g$.}\label{fig:06}
\end{figure}

To analyze the effect of $\We_g$ on the droplet deformation, the variations of the droplet width and the droplet thickness with $\We_g$ are shown in Figure \ref{fig:06}. The deformation characteristics at the moment when the aspect ratio ($d_w / d_t$) is its largest is extracted, i.e., the peak moment in Figure \ref{fig:05}b. As $\We_g$ increases, the droplet width ($d_w$) first increases and then decreases, as shown in Figure \ref{fig:06}a. The first increase is due to the increased droplet deformation caused by the enhanced aerodynamic forces. Then, as $\We_g$ increases further, the droplet periphery may be deflected towards the leeward of the droplet \cite{Xu2023Viscosity}, as shown in Figure \ref{fig:06}c. The deflection of the droplet periphery has two effects on the width ($d_w$) and thickness ($d_t$) of the droplet. First, the deflection of the droplet periphery itself slows down the increment of $d_w$ or even causes it to decrease. Second, the deflection causes $d_t$ to increase rapidly, leading to an earlier moment when ${d_w}/{d_t}$ is its largest (i.e., measured at an earlier moment). These effects cause the later decrease in $d_w$ at a high $\We_g$. To further compare the droplet width, we calculated the average droplet width, i.e., the dashed lines in Figure \ref{fig:06}a. The average width of the compound droplets is smaller than that of the single-component droplets and decreases as the volume ratio of the water core increases. This is because the water core is less deformed and contributes less to the overall width of the compound droplet.

In contrast to the width, the thickness of the droplet first decreases and then increases as $\We_g$ increases, as shown in Figure \ref{fig:06}b. Under a low $\We_g$, the thickness of the compound droplet mainly depends on the thickness of the less deformed water core, while the single-component silicone oil droplet is completely flattened, so the thickness of the compound droplet is larger than that of the single-component droplet. However, when $\We_g$ reaches a certain value (about 45), the thickness of the compound droplet and the single-component droplet are similar. This is because the periphery of the single-component droplet deforms and deflects rapidly at a high $\We_g$, leading to less flattening of the droplet middle when the aspect ratio is its largest. The less flattening of the single-component droplet and the less deformation of the water core of the compound droplet produces a similar effect. In addition, as shown in Figure \ref{fig:06}b, the thickness of the compound droplet is almost independent of the volume ratio ($\alpha $). This is because the change of the water core size with $\alpha $ is small and the water cores are flattened more at a larger $\alpha $, which leads to a weak dependence of the compound droplet thickness (determined by the thickness of the water core) on $\alpha $.

\subsection{Regime map of compound droplet breakup}\label{sec:33}
\begin{figure}
  \centering
  \includegraphics[width=0.55\columnwidth]{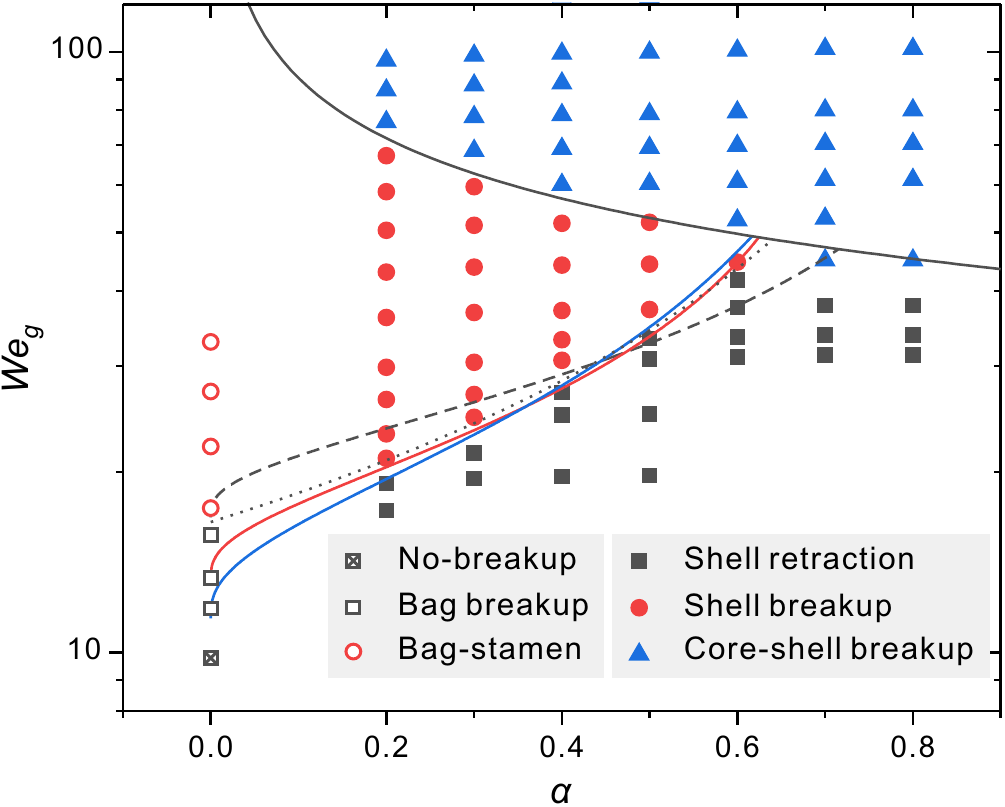}
  \caption{Regime map of droplet breakup. The solid symbols correspond to the breakup modes of compound droplets, and the hollow symbols correspond to the breakup modes of single-component droplets. The black solid line corresponds to Eq.\ (\ref{eq:eq01}) for the transition between the shell and core-shell breakup. For the transition between the shell retraction and the shell breakup, the dashed line is Eq.\ (\ref{eq:eq10}), the red solid line is Eq.\ (\ref{eq:eq13}), the blue solid line is Eq.\ (\ref{eq:eq16}), and the dotted line is Eq.\ (\ref{eq:eq17}).}\label{fig:07}
\end{figure}

The breakup of compound droplets is classified into shell retraction, shell breakup, and core-shell breakup according to the droplet morphologies, as described in Section \ref{sec:31}. In this section, a regime map of these breakup modes (Figure \ref{fig:07}) is obtained in the parameter space of $\We_g$ and $\alpha$. The transition conditions between different breakup modes are discussed in this section.

The criterion between the shell breakup and the core-shell breakup is whether the water core breaks up. Similar to the breakup of the single-component droplet, whether the water core breaks up depends on the competition between aerodynamic force and the surface tension force of the water core, i.e., the Weber number of the water core (${{\We}_{g,wc}}$) reaches a critical value. Since the sum of the oil surface tension and the oil-water interfacial tension is smaller than the water surface tension (${\sigma _o} + {\sigma _{ow}} < {\sigma _w}$), the water core remains wetted by oil, so the sum is used in the definition of ${{\We}_{g,wc}}$, and  ${{\We}_{g,wc}}={{{\rho }_{g}}u_{g}^{2}{{d}_{c}}}/{\left( {{\sigma }_{o}}+{{\sigma }_{ow}} \right)}$. ${{\We}_{g,wc}}$ can be converted from $\We_g$ by relating the diameters of the compound droplet ($d_0$) and the water core (${{d}_{c}}={{d}_{0}}{{\alpha }^{1/3}}$) and replacing the surface tension of the compound droplet ($\sigma_o$) with that of the water core (${{\sigma }_{o}}+{{\sigma }_{ow}}$). Therefore, this transition corresponds to
\begin{equation}\label{eq:eq01}
  {{ \We}_{g,wc}}={{\We}_{g}}{{\alpha }^{1/3}}\frac{{{\sigma }_{o}}}{{{\sigma }_{o}}+{{\sigma }_{ow}}}={{C}_{1}}
\end{equation}
The relation is shown as the solid line between the shell breakup and core-shell breakup regimes in Figure \ref{fig:07}, where $C_1 = 14$ is obtained by fitting the critical points between the shell breakup and the core-shell breakup. We further discuss the applicability of Eq.\ (\ref{eq:eq01}) to the special case of a single-component droplet breakup. If $\alpha $ = 0, the droplet consists entirely of oil and there is no water core. If $\alpha $ is close to 0, the water core is very small, so a very large Weber number of the compound droplet (${\We_g}$) is needed to maintain ${\We_{g,wc}} = {C_1}$, corresponding to ${\We_g} \to \infty $ at $\alpha \to 0$ in Figure \ref{fig:07}. If $\alpha $ is close to 1, the droplet consists mainly of water, and the transition of whether the water core breaks up corresponds to ${\We_{g,wc}}$ = 14, which is close to the critical Weber number (11 ± 2) of whether single-component droplets break up or not (i.e., the critical Weber number of single-component droplets between the no-breakup and the bag breakup mode) \cite{Guildenbecher2009SecondaryAtomization, Jain2015SecondaryBreakup, Yang2017TransitionsWeberNumber}. Here, ${{\We}_{g,wc}}$ is slightly larger. This is because the tension of the peripheral oil shell on the water core causes a larger loss of the relative velocity between the airflow and the water core after the initial flattening of the water core, so a larger airflow velocity (i.e., a larger ${{\We}_{g,wc}}$) is required to pierce the water core.

\begin{figure}
  \centering
  \includegraphics[width=0.4\columnwidth]{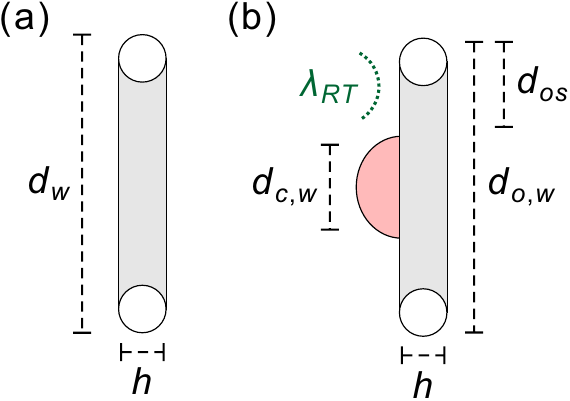}
  \caption{Illustration of droplet geometry after the initial flattening of the droplet: (a) single-component droplet, (b) compound droplet.}\label{fig:08}
\end{figure}

The transition between the shell retraction and the shell breakup depends on whether the oil shell is pierced by the airflow. Previous studies have indicated that the piercing is controlled by the development of Rayleigh-Taylor (RT) instabilities on the windward side after the droplet is flattened \cite{Guildenbecher2009SecondaryAtomization, Sharma2022BreakupReview, Theofanous2011DropBreakup, Zhao2011BagBreakup}. For the single-component droplet, the windward side of the droplet is flattened into a disk (as shown in Figure \ref{fig:08}a), so whether the droplet can be pierced by the airflow is determined by the comparison between the most-amplified wavelength of the RT instability ($\lambda_{RT}$) and the droplet width after the initial flattening ($d_w$) \cite{Xu2023Viscosity}. However, for the compound droplet, the windward side is disrupted by the less deformed water core, as shown in Figure \ref{fig:08}b. With this morphology, whether the peripheral oil shell is pierced depends on the comparison between the width of the oil shell on one side ($d_{os}$) and $\lambda_{RT}$. That is, the transition between the shell retraction and the shell breakup corresponds to
\begin{equation}\label{eq:eq02}
  \frac{{{d}_{os}}}{{{\lambda }_{RT}}}=C
\end{equation}
where $C$ is a constant. According to the classical dispersion relation of the RT instability where an infinite planar surface is assumed \cite{Chandrasekhar2013Instability}, the most-amplified wavelength $\lambda_{RT}$ in the inviscid case is
\begin{equation}\label{eq:eq03}
  {{\lambda }_{RT}}=2\pi \sqrt{\frac{3{{\sigma }_{o}}}{a\left( {{\rho }_{o}}-{{\rho }_{g}} \right)}}
\end{equation}
where ${{\sigma }_{o}}$ and ${{\rho }_{o}}$ are the surface tension and the density of the silicone oil, and ${{\rho }_{g}}$ is the density of the airflow and ${{\rho }_{g}}\ll {{\rho }_{o}}$. This most-amplified wavelength is an approximation of the actual scenario, but provides a suitable length scale to characterize the RT instability wavelength. $a$ is the acceleration of the peripheral oil shell after the initial flattening and can be considered as
\begin{equation}\label{eq:eq04}
  a=\frac{{{F}_{D}}}{m}=\frac{{{C}_{D}}{{\rho }_{g}}u_{g}^{2}}{2{{\rho }_{o}}h}
\end{equation}
where ${{F}_{D}}$ is the airflow drag force and ${{F}_{D}}={{C}_{D}}{{\rho }_{g}}u_{g}^{2}{{A}_{p}}/2$, $m$ is the mass of the oil shell and $m={{\rho }_{o}}h{{A}_{p}}$, ${{C}_{D}}$ is the drag coefficient and considered as a constant $C_D = 1.2$ for a disk \cite{White2003FluidMechanics}, ${{A}_{p}}$ and $h$ are the projected area and the thickness of the peripheral oil shell after the initial flattening, respectively. According to the pressure balance at the rim of the oil shell between the capillary pressure ($\sim 2{{\sigma }_{o}}/h$) and the airflow pressure ($\sim {{\rho }_{g}}u_{g}^{2}/2$), the thickness of the peripheral oil shell can be estimated as
\begin{equation}\label{eq:eq05}
  \frac{h}{{{d}_{0}}}\sim \frac{4}{ {{ \We}_{g}}}
\end{equation}
Combining Eqs.\ (\ref{eq:eq03})--(\ref{eq:eq05}), we can obtain
\begin{equation}\label{eq:eq06}
  \frac{{{\lambda }_{RT}}}{{{d}_{0}}}\sim \frac{4\sqrt{5}\pi }{ {{ \We}_{g}}}
\end{equation}
Moreover, according to the droplet geometry after the initial flattening of the oil shell (as illustrated in Figure \ref{fig:08}b), $d_{os}$ is equal to
\begin{equation}\label{eq:eq07}
  \frac{{{d}_{os}}}{{{d}_{0}}}=\frac{{{d}_{o,w}}-{{d}_{c,w}}}{2{{d}_{0}}}
\end{equation}
where ${{d}_{o,w}}$ is the disk width of the oil shell, and ${{d}_{c,w}}$ is the width of the water core. Considering that the disk formed by the flattening of the oil shell overlaps less with the water core, we assume that the liquid volume of the disk is equal to the volume of the oil. Therefore, based on this assumption and Eq.\ (\ref{eq:eq05}), we can simplify the disk width of the oil shell (${{d}_{o,w}}$) as
\begin{equation}\label{eq:eq08}
  \frac{{{d}_{o,w}}}{{{d}_{0}}}=\sqrt{\frac{2{{d}_{0}}\left( 1-\alpha  \right)}{3h}}\sim \sqrt{\frac{ {{ \We}_{g}}\left( 1-\alpha  \right)}{6}}
\end{equation}
In addition, at the transition between the shell retraction and the shell breakup, the Weber number of the water core (${{\We}_{g,wc}}$) is small due to the larger surface tension and the smaller core size. The range of ${{\We}_{g,wc}}$ is about 3--12. If we neglect the deformation of the water core at these $ {{ \We}_{g,wc}}$, i.e., the width of the water core (${{d}_{c,w}}$) is equal to the diameter of the water core ($d_c$), ${{d}_{c,w}}$ can be estimated as
\begin{equation}\label{eq:eq09}
  \frac{{{d}_{c,w}}}{{{d}_{0}}}={{\alpha }^{1/3}}
\end{equation}
Finally, combining Eqs.\ (\ref{eq:eq02}), (\ref{eq:eq06}), and (\ref{eq:eq07}), we can obtain the following model
\begin{equation}\label{eq:eq10}
  \frac{\We_{g}^{3/2}{{\left( 1-\alpha  \right)}^{1/2}}-\sqrt{6} {{ \We}_{g}}{{\alpha }^{1/3}}}{8\sqrt{30}\pi }={{C}_{2}}
\end{equation}
The model of Eq.\ (\ref{eq:eq10}) is shown as the dashed line between the shell retraction regime and the shell breakup regime in Figure \ref{fig:07}, where ${C_2}$ = 0.5 is obtained by fitting the critical points between the shell breakup and the shell retraction. Compared with the experimental data for the transition between the shell retraction and the shell breakup, the dependence of the critical $\We_g$ on $\alpha$ in the model of Eq.\ (\ref{eq:eq10}) is slightly weaker.

The above analysis neglects the deformation of the water core. To further improve the model, we now consider the deformation of the water core in the model. Since the water core is surrounded by an oil shell, it is difficult to measure its degree of deformation directly. However, for the small $ {{ \We}_{g,wc}}$, the deformation of the water core can be estimated through the Taylor analogy breakup (TAB) model, which describes the droplet dynamics by considering the analogy with a damped spring-mass system \cite{Hoang2019TAB, Amsden1987TAB, Rimbert2020DropletDeformation}. In the inviscid case of the TAB model, the maximum width of the water core, which corresponds to the maximum amplitude in a damped spring-mass system, can be estimated as
\begin{equation}\label{eq:eq11}
  \frac{{{d}_{c,w}}}{{{d}_{c}}}=1+\frac{ {{ \We}_{g,wc}}}{24}
\end{equation}
where $ {{ \We}_{g,wc}}$ is the Weber number of the water core and can be converted from $\We_g$ through Eq.\ (\ref{eq:eq01}). Therefore, if we consider the deformation of the water core through Eq.\ (\ref{eq:eq11}), we can replace Eq.\ (\ref{eq:eq09}) with
\begin{equation}\label{eq:eq12}
  \frac{{{d}_{c,w}}}{{{d}_{0}}}=\left( 1+\frac{ {{ \We}_{g}}{{\alpha }^{1/3}}}{72} \right){{\alpha }^{1/3}}
\end{equation}
Then, Eq.\ (\ref{eq:eq10}) is converted to
\begin{equation}\label{eq:eq13}
  \frac{\We_{g}^{3/2}{{\left( 1-\alpha  \right)}^{1/2}}-\sqrt{6}\left( 1+\frac{ {{ \We}_{g}}{{\alpha }^{1/3}}}{72} \right) {{ \We}_{g}}{{\alpha }^{1/3}}}{8\sqrt{30}\pi }={{C}_{3}}
\end{equation}
The model of Eq.\ (\ref{eq:eq13}) is shown as the red solid line between the shell retraction regime and the shell breakup regime in Figure \ref{fig:07}, where ${C_3}$ = 0.35 is obtained by fitting the critical points between the shell breakup and the shell retraction. The transition between the shell retraction and the shell breakup can be well predicted by this model.

In addition, Hsiang and Faeth \cite{Hsiang1992SecondaryBreakup} proposed another semi-empirical formula to quantify droplet deformation. Based on the parameters of the water core deformation in this study, we rewrite their semi-empirical formula as
\begin{equation}\label{eq:eq14}
  \frac{{{d}_{c,w}}}{{{d}_{c}}}=1+0.19\We_{g,wc}^{1/2},\quad {{ \We}_{g,wc}}<100,\Oh<0.1
\end{equation}
Therefore, if we consider the deformation of the water core through Eq.\ (\ref{eq:eq14}), we can replace Eq.\ (\ref{eq:eq09}) with
\begin{equation}\label{eq:eq15}
  \frac{{{d}_{c,w}}}{{{d}_{0}}}=\left( 1+0.11\We_{g}^{1/2}{{\alpha }^{1/6}} \right){{\alpha }^{1/3}}
\end{equation}
Using Eq.\ (\ref{eq:eq15}), we can convert Eq.\ (\ref{eq:eq10}) to
\begin{equation}\label{eq:eq16}
  \frac{\We_{g}^{3/2}{{\left( 1-\alpha  \right)}^{1/2}}-\sqrt{6}\left( 1+0.11\We_{g}^{1/2}{{\alpha }^{1/6}} \right) {{ \We}_{g}}{{\alpha }^{1/3}}}{8\sqrt{30}\pi }={{C}_{4}}
\end{equation}
The model of Eq.\ (\ref{eq:eq16}) is shown as the blue solid line between the shell retraction regime and the shell breakup regime in Figure \ref{fig:07}, where ${C_4}$ = 0.28 is obtained by fitting the critical points between the shell breakup and the shell retraction.

The applicability of Eqs.\ (\ref{eq:eq10}), (\ref{eq:eq13}), and (\ref{eq:eq16}) to the special case of a single-component droplet breakup can be further discussed. The special case of Eqs.\ (\ref{eq:eq10}), (\ref{eq:eq13}), and (\ref{eq:eq16}) for the single-component droplet case correspond to the case where $\alpha $ = 0, i.e., the droplet consists entirely of oil. Since Eqs.\ (\ref{eq:eq10}), (\ref{eq:eq13}), and (\ref{eq:eq16}) correspond to the criterion of whether one side of the droplet can be pierced, the Weber numbers $\We_g$ obtained from Eqs.\ (\ref{eq:eq10}), (\ref{eq:eq13}), and (\ref{eq:eq16}) in the case of $\alpha $ = 0 should correspond to the critical Weber number between the bag breakup and the bag-stamen breakup of a single-component droplet. As shown by the hollow symbols in Figure \ref{fig:07}, the critical Weber number between the bag breakup and the bag-stamen breakup of a single-component oil droplet is about 16.5 in our experiment. Correspondingly, the Weber numbers ($\We_g$) obtained from Eqs.\ (\ref{eq:eq10}), (\ref{eq:eq13}), and (\ref{eq:eq16}) are 16.8, 13.24, and 11.41 in the case of $\alpha $ = 0. Comparing the critical Weber number of single-component oil droplets (about 16.5), the Weber numbers ($\We_g$) obtained from Eqs.\ (\ref{eq:eq10}), (\ref{eq:eq13}), and (\ref{eq:eq16}) in the case of $\alpha $ = 0 is within a reasonable range due to some estimation in the modeling process. These estimations include the estimations for sizes of the droplet shape and the neglect of the effect of finite interface on RT instability.

In addition, the models of Eqs.\ (\ref{eq:eq10}), (\ref{eq:eq13}), and (\ref{eq:eq16}) may be too complex for practical use. Considering that the transition Weber number in the case of $\alpha $ = 0 should be the critical Weber number between the bag breakup and the bag-stamen breakup of single-component oil droplets (about 16.5 in our experiments), we also give a correlation by direct fitting as
\begin{equation}\label{eq:eq17}
   {{\We_g} = 16.5{\left( {1 - \alpha } \right)^{1.06}}}
\end{equation}
where the exponent index is obtained by fitting the critical points between the shell breakup and the shell retraction. The correlation of Eq.\ (\ref{eq:eq17}) is shown as the dotted line in Figure \ref{fig:07}.

\subsection{Effect of water core eccentricity on droplet breakup}\label{sec:34}
In the previous sections, we consider cases when the eccentricity of the water core is zero. To further explore the effect of the eccentricity, we increase the droplet falling height ($H$) so that the core moves to the bottom of the shell before entering the airflow due to the water-oil density difference, corresponding to the case of the maximum eccentricity.

\begin{figure}
  \centering
  \includegraphics[width=0.7\columnwidth]{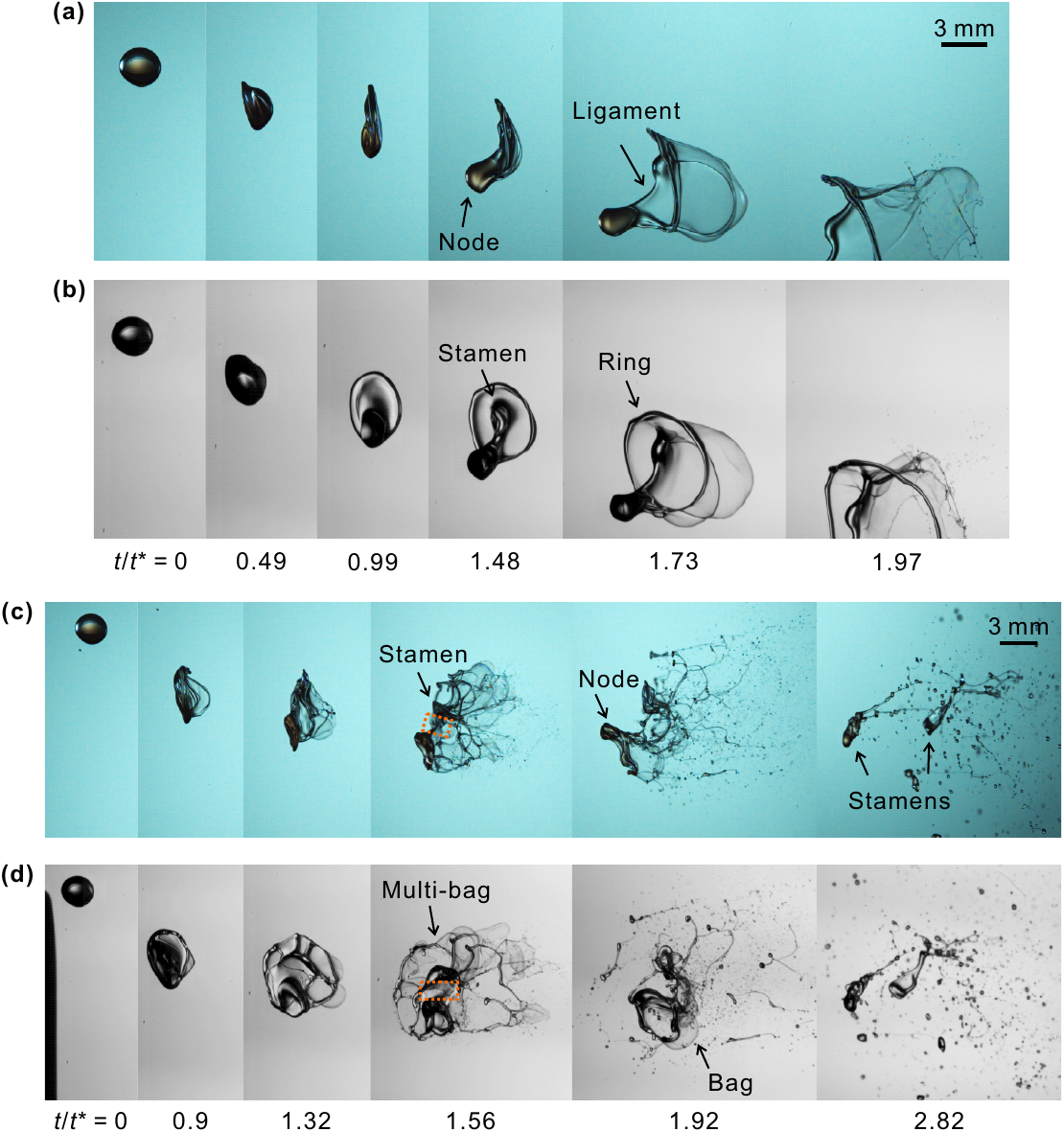}
  \caption{(a-b) Shell breakup of a compound droplet with a maximum eccentricity, (a) side view, (b) 45$^\circ$ view. $\We_g = 30.4$, $\alpha = 0.3$, $H = 600$ mm. (c-d) Core-shell breakup of a compound droplet with a maximum eccentricity, (c) side view, (d) 45$^\circ$ view. $\We_g = 77.9$, $\alpha = 0.3$, $H = 600$ mm. The whole processes are shown as Videos 5-6 in Supplementary Material.}\label{fig:09}
\end{figure}

Figures \ref{fig:09}a-b show the shell breakup process of a compound droplet with a maximum eccentricity, which has the same $\We_g$ and $\alpha$ as Figure \ref{fig:03}a. Different from Figure \ref{fig:03}a, the water core in Figures \ref{fig:09}a-b is at the bottom of the oil shell and forms a large node at the bottom of the compound droplet after the flattening. Meanwhile, the oil shell forms a bag-stamen structure ($t/t^* = 1.48$ in Figures \ref{fig:09}a-b). The bottom node formed by the water core and the middle stamen formed by the oil shell are connected by a thick ligament. Between the thick ligament and the peripheral ring of the oil shell, bags develop ($t/t^* = 1.73$ in Figures \ref{fig:09}a-b). As the bags develop, the ligament and the peripheral ring are gradually stretched ($t/t^* =$ 1.73--1.97 in Figures \ref{fig:09}a-b).

Figures \ref{fig:09}c-d show the core-shell breakup process of a compound droplet with a maximum eccentricity. The oil shell forms a multi-bag structure with a stamen in the middle ($t/t^* = 1.56$ in Figures \ref{fig:09}c-d), while the water core forms a bag structure ($t/t^* = 1.92$ in Figures \ref{fig:09}c-d). The liquid between the stamen in the oil shell middle and the water core (highlighted by the orange dashed rectangle) deforms more slowly than the liquid around the oil shell, but it can still be pierced in the end ($t/t^* = 1.92$ in Figures \ref{fig:09}c-d). The piercing of this part of the liquid can tense the upper part of the water core, causing a large node to form in the upper part of the water core, which further develops into a stamen. Finally, two stamens are formed, one composed mainly of water and the other of oil ($t/t^* = 2.82$ in Figures \ref{fig:09}c-d).

\section{Conclusions}\label{sec:4}
We experimentally study the deformation and breakup of compound droplets in airflows. The compound droplet consists of a water core and a silicone oil shell and is produced by a coaxial needle method \cite{Nathan2020CompoundDropletRebound, Blanken2021CompoundDropletImpact}. Different from the single-component droplets \cite{Sharma2022BreakupReview, Theofanous2011DropBreakup, Xu2023Viscosity}, the compound droplets exhibit different deformation and breakup morphologies due to the different deformation rates of different components in the airflow. According to whether and where the breakup occurs, we classify the compound droplet breakup into shell retraction (where the oil shell is flattened and then retracts), shell breakup (where the oil shell breaks up), and core-shell breakup (where both the oil shell and the water core break up).

The evolution of deformation characteristics of the compound droplet with time and $\We_g$ is analyzed quantitatively. Compared with the single-component droplet \cite{Guildenbecher2009SecondaryAtomization}, the compound droplet stretches more in the flow direction and expands less in the cross-flow direction due to the slower deformation of the water core than that of the oil shell. The velocity analysis shows the deformation difference between compound and single-component droplets occurs when the water core starts to protrude into the airflow. A map of breakup modes of compound droplets is obtained in the parameter space of $\We_g$ and $\alpha$. The transition between the shell retraction and the shell breakup depends on whether the oil shell is pierced by the airflow, and the transition between the shell and core-shell breakup depends on whether the water core breaks up. Based on the corresponding transition condition, the transitions are modeled theoretically. Differences in the windward morphology and the deformation rate lead to different transition models for compound droplets than for single-component droplets \cite{Xu2023Viscosity, Zhao2011BagBreakup}. Finally, the effects of eccentricity are discussed. When the eccentricity is maximum, the liquid between the core and the middle of the shell deforms more slowly than the other part of the shell, which causes the droplet to form new structures, such as thick ligaments or two stamens.

The breakup of compound droplets is a complex process of interfacial evolution because of the liquid-liquid interface. This study classifies the breakup of compound droplets into several main modes based on whether and where the breakup occurs. More sub-modes can be extended based on more detailed interfacial morphology, like different breakup structures \cite{Guildenbecher2009SecondaryAtomization, Sharma2022BreakupReview} of the core or the shell. Especially, the Kelvin-Helmholtz instability appears at the droplet edge when the airflow velocity is increased to a certain extent \cite{Theofanous2011DropBreakup}. The effect of the liquid-liquid interface in the compound droplet in this condition can be further explored.
In addition, the droplet viscosity can affect the breakup of single-component droplets \cite{Theofanous2012ViscousLiquids, Xu2023Viscosity}, and its effect on compound droplets can be further studied in the future.


\section*{Acknowledgements}
This work is supported by the National Natural Science Foundation of China (Grant Nos.\ 51676137, 52176083, and 51920105010).


\bibliography{CompoundDropletBreakup}
\end{document}